\documentstyle [12pt,twoside]{article}

\newcommand{\bq}{\begin{equation}}
\newcommand{\ba}{\begin{eqnarray}}
\newcommand{\eq}{\end{equation}}
\newcommand{\ea}{\end{eqnarray}}

\def\b{\beta}

\def\d{\delta}
\def\e{\epsilon}
\def\f{\phi}

\def\l{\lambda}

\def\p{\pi}

\def\s{\sigma}

\def\D{\Delta}
\def\F{\Phi}
\def\G{\Gamma}
\def\J{\Psi}
\def\L{\Lambda}
\def\O{\Omega}

\def\bo{{\raise.15ex\hbox{\large$\Box$}}}
\def\bob{{\lower.2ex\hbox{\large$\Box$}}}
\def\pa{\partial}
\def\de{\nabla}

\def\TH{{\raise.2ex\hbox{$\displaystyle \bigodot$}\mskip-4.7mu \llap H \;}}
\def\face{{\raise.2ex\hbox{$\displaystyle \bigodot$}\mskip-2.2mu \llap {$\ddot
        \smile$}}}

\def\Hat#1{\rlap{\kern.10em$\widehat{\phantom G}$}#1}
\def\HAt#1{\rlap{\kern.05em$\widehat{\phantom G}$}#1}

\def\cap#1{\rlap{\kern.1em$\widehat{\phantom{G\vrule height.8em}}$}#1{}}
\def\Cap#1{\rlap{\kern.05em$\widehat{\phantom{G\vrule height.8em}}$}#1{}}
\def\bra#1{\left\langle #1\right|}
\def\ket#1{\left| #1\right\rangle}

\def\VEV#1{\left\langle #1\right\rangle}

\def\abs#1{\left| #1\right|}
\def\leftrightarrowfill{$\mathsurround=0pt \mathord\leftarrow \mkern-6mu
        \cleaders\hbox{$\mkern-2mu \mathord- \mkern-2mu$}\hfill
        \mkern-6mu \mathord\rightarrow$}
\def\overleftrightarrow#1{\vbox{\ialign{##\crcr
        \leftrightarrowfill\crcr\noalign{\kern-1pt\nointerlineskip}
        $\hfil\displaystyle{#1}\hfil$\crcr}}}

\def\frac#1#2{{\textstyle{#1\over\vphantom2\smash{\raise.20ex
        \hbox{$\scriptstyle{#2}$}}}}}

\catcode`@=11
\def\underline#1{\relax\ifmmode\@@underline#1\else
        $\@@underline{\hbox{#1}}$\relax\fi}
\catcode`@=12

\def\nis{\nointerlineskip}
\def\Abar{\vbox{\nis\moveright.33em\vbox{
        \hrule width.35em height.04em}\nis\kern.05em\hbox{$A$}}{}}
\def\Dbar{\vbox{\nis\moveright.20em\vbox{
        \hrule width.50em height.04em}\nis\kern.05em\hbox{$D$}}{}}
\def\Gbar{\vbox{\nis\moveright.20em\vbox{
        \hrule width.50em height.04em}\nis\kern.05em\hbox{$G$}}{}}
\def\mbar{\vbox{\nis\moveright.15em\vbox{
        \hrule width.60em height.04em}\nis\kern.05em\hbox{$m$}}{}}
\def\Rbar{\vbox{\nis\moveright.20em\vbox{
        \hrule width.50em height.04em}\nis\kern.05em\hbox{$R$}}{}}
\def\Vbar{\vbox{\nis\moveright.05em\vbox{
        \hrule width.60em height.04em}\nis\kern.05em\hbox{$V$}}{}}
\def\Xbar{\vbox{\nis\moveright.20em\vbox{
        \hrule width.60em height.04em}\nis\kern.05em\hbox{$X$}}{}}
\def\thetabar{\vbox{\nis\moveright.15em\vbox{
        \hrule width.30em height.04em}\nis\kern.05em\hbox{$\theta$}}{}}
\def\Lambdabar{\vbox{\nis\moveright.25em\vbox{
        \hrule width.35em height.04em}\nis\kern.05em\hbox{${\mit\Lambda}$}}{}}
\def\Sigmabar{\vbox{\nis\moveright.25em\vbox{
        \hrule width.50em height.04em}\nis\kern.05em\hbox{${\mit\Sigma}$}}{}}
\def\phibar{\vbox{\nis\moveright.18em\vbox{
        \hrule width.40em height.04em}\nis\kern.05em\hbox{$\phi$}}{}}
\def\chibar{\vbox{\nis\moveright.12em\vbox{
        \hrule width.40em height.04em}\nis\kern.05em\hbox{$\chi$}}{}}
\def\psibar{\vbox{\nis\moveright.23em\vbox{
        \hrule width.40em height.04em}\nis\kern.05em\hbox{$\psi$}}{}}
\def\debar{\vbox{\nis\moveright.18em\vbox{
        \hrule width.35em height.04em}\nis\kern.05em\hbox{$\partial$}}{}}
\def\delbar{\vbox{\nis\moveright.10em\vbox{
        \hrule width.63em height.04em}\nis\kern.05em\hbox{$\nabla$}}{}}

\begin{document}

\centerline{\large{\bf{Statistical Mechanics of Kinks in
(1+1)-Dimensions:}}}
\centerline{\large{\bf{Numerical Simulations and Double Gaussian
Approximation}}}

\vspace{1cm}

\centerline{Francis J. Alexander,$^{\star}$ Salman Habib,$^{\dagger}$
and Alex Kovner$^*$}

\vspace{.3cm}

\centerline{\em $^{\star}$Center for Nonlinear Studies}
\centerline{\em Los Alamos National Laboratory}
\centerline{\em Los Alamos, NM 87545}

\vspace{.3cm}

\centerline{\em $^{\dagger}$T-6, Theoretical Division}
\centerline{\em Los Alamos National Laboratory}
\centerline{\em Los Alamos, NM 87545}

\vspace{.3cm}

\centerline{\em $^*$T-8, Theoretical Division}
\centerline{\em Los Alamos National Laboratory}
\centerline{\em Los Alamos, NM 87545}

\vspace{.7cm}

\centerline{\bf Abstract}

{\small We investigate the thermal equilibrium properties of kinks
in a classical $\F^4$ field theory in $1+1$ dimensions. From large
scale Langevin simulations we identify the temperature below which a
dilute gas description of kinks is valid. The standard dilute gas/WKB
description is shown to be remarkably accurate below this temperature.
At higher, ``intermediate'' temperatures, where kinks still exist, this
description breaks down. By introducing a double Gaussian variational
ansatz for the eigenfunctions of the statistical transfer operator for
the system, we are able to study this region analytically. In
particular, our predictions for the number of kinks and the
correlation length are in agreement with the simulations. The double
Gaussian prediction for the characteristic temperature at which the
kink description ultimately breaks down is also in accord with the
simulations. We also analytically calculate the internal energy and
demonstrate that the peak in the specific heat near the kink
characteristic temperature is indeed due to kinks. In the neighborhood
of this temperature there appears to be an intricate energy
sharing mechanism operating between nonlinear phonons and kinks.}

\vfill
\noindent e-mail:\\
\noindent fja@goshawk.lanl.gov\\
\noindent habib@eagle.lanl.gov\\
\noindent kovner@pion.lanl.gov\\
\newpage

\centerline{\bf I. Introduction}

The equilibrium and nonequilibrium statistical mechanics of solitons,
solitary waves, and other coherent structures in nonlinear systems has
been a subject of study for some time \cite{SS}. Recent interest
has been fueled by new applications not only in condensed matter
physics \cite{CM}\cite{BKT}, but also by potential applications in
particle physics (sphalerons) \cite{SR} and cosmology  (domain walls,
baryogenesis) \cite{KT}.

In this paper we focus on the classical equilibrium statistical
mechanics of solitary wave (``kink'') solutions of a tachyonic mass,
$\phi^4$ (``double well'') field theory in $1+1$ spacetime dimensions
with the Lagrangian density
\begin{eqnarray}
L= {1\over 2}\left(\partial_t\Phi\right)^2
   -{1\over 2}\left(\partial_x\Phi\right)^2
   +{1\over 2}m^2\Phi^2-{1\over 4}\L\Phi^4,       \label{1}
\end{eqnarray}
and the corresponding equation of motion
\bq
\pa_{tt}^2\F=\pa_{xx}^2\F+m^2\F-\L\F^3.                    \label{2}
\eq
This model is of direct relevance to the study of displacive
phase transitions \cite{ADB2} and magnetic spin chains \cite{MS1}.
Moreover, the behavior of this model is representative of a large
class of soliton-bearing systems.  Finally, it has the added advantage
of being amenable to both theoretical analysis and numerical
simulation.

The kink solution of the equation of motion (\ref{2}) is the localized
field configuration that interpolates between the two asymptotic field
potential energy minima at $\F=\pm \F_0=\pm m/\sqrt{\L}$. Kink
solutions of the field equations are inaccessible to
perturbation theory: A statistical mechanics of kinks is nevertheless
still possible, partly because they are localized objects. The static
kink solution centered at $x=x_0$ is
\bq
\F_k= {m\over\sqrt{\L}}\hbox{tanh}\left({m\over\sqrt{2}}(x-x_0)\right),
                                              \label{3}
\eq
and the negative of the kink solution is the anti-kink. Time dependent
solutions are easily found by Lorentz boosting the static solution.
The energy density of the static kink (or anti-kink) is
\bq
\e_k(x)={m^4\over 2\L}\hbox{sech}^4\left({m(x-x_0)\over\sqrt{2}}\right),
                                             \label{4}
\eq
from which the total energy of an isolated kink, or the kink mass, is
obtained:
\ba
E_k&=&\int_{-\infty}^{+\infty}dx~\e_k(x)                  \label{5}\\
   &=& \sqrt{8\over 9}{m^3\over\L}.                          \label{6}
\ea

In numerical simulations it is customary to employ a dimensionless
form of the theory given by the transformations:
\ba
\f&=&\F /a,                      \nonumber\\
\bar{x}&=&mx,                          \nonumber\\
\bar{t}&=&mt,                                \label{7}
\ea
where $a^2=m^2/\L$. Applying these transformations, we have
\bq
H=ma^2\bar{H}                         \label{8}
\eq
where $H$ is the original field Hamiltonian, and
\bq
\bar{H}=\int d\bar{x}~\left[{1\over 2}\bar{\p}^2+{1\over
2}\left(\bar{\de}{\f}\right)^2\-{1\over 2}{\f}^2 +{1\over
4}{\f}^4\right].    \label{9}
\eq
The new field $\f$ satisfies the equation of motion
\bq
\pa_{\bar{t}\bar{t}}^2\f=\pa_{\bar{x}\bar{x}}^2\f-\f(\f^2-1).
                                                  \label{10}
\eq
We enforce $\bar{\b}\bar{H}=\b H$ by introducing a new scaled
temperature $\bar{\b}=\b/(ma^2)$.

The statistical mechanics of kinks in this system has been studied by
two approaches. In the first, and phenomenological, approach one assumes
that the kinks and the field fluctuations about the asymptotic field
minima (``phonons'') may be treated as weakly interacting elementary
excitations. Provided that the kink density is low (the dilute gas
approximation), the canonical partition function can be found by
standard methods \cite{SS}\cite{KS}\cite{CKBT}. Alternatively, as
shown by Krumhansl and Schrieffer (KS) \cite{KS}, building on earlier
work of Sears, Scalapino, and Ferrell \cite{SSF}, it is possible to
calculate the partition function, in principle exactly, by exploiting
a transfer operator technique. KS showed that in the low temperature
(dilute gas) limit the partition function naturally factorizes into
two contributions both having counterparts in the phenomenological
theory; a tunneling term which they were able to identify with the
kink contribution, and the remainder which they identified as the
linearized phonons.  The approach of KS was further refined and
extended to a wider class of systems by Currie, Krumhansl, Bishop, and
Trullinger \cite{CKBT}. In this work, interactions of kinks with
linearized phonons were taken into account, leading to substantial
corrections to the results of KS.

The key result of these efforts is the prediction that, below a certain
temperature, the spatial density of kinks is
\bq
n_k\propto\sqrt{E_k\b}\exp(-E_k\b).                    \label{11}
\eq
A related prediction is that the field correlation length $\l$
defined by
\bq
\VEV{\F(0)\F(x)}\sim\exp(-\abs{x}/\l)                      \label{12}
\eq
exhibits an exponential temperature dependence \cite{CKBT},
\bq
\l={1\over 4}\sqrt{\p\over 3}{1\over\sqrt{E_k\b}}\exp(E_k\b)
                                                   \label{13}
\eq
at low temperatures.

Computer simulations to verify these results date back to Ref.
\cite{KBKS} where only a qualitative agreement was found. Recent work
\cite{GR}\cite{BD}\cite{AFG} has led to more detailed comparisons,
however significant discrepancies have been reported. This has led to
theoretical speculation \cite{FM}\cite{BF2} regarding possible
corrections to the dilute gas theory of kinks. It has been suggested
that these discrepancies are due to finite size effects and phonon
dressing of the bare kink energy (breather contributions to the free
energy may also be significant \cite{SSB}). As was discussed in
Ref. \cite{AH}, the earlier simulations were not carried out at low enough
temperatures: Nevertheless the authors interpreted their results in terms
of WKB formulae that are simply not a valid description over the range of
temperatures they had studied. In this intermediate temperature
regime, there is no unique characterization of what constitutes a kink
and how to differentiate it from a nonlinear phonon. By going to low
enough temperatures where the dilute gas results are valid, we are
also able to rule out an earlier claim of substantial phonon dressing
even at these temperatures \cite{KBKS}.

We have numerically studied the equilibrium statistical mechanics of
kinks in the $\Phi^4$ model by implementing a Langevin code on a
massively parallel computer. To understand our results in the high
and intermediate temperature region not susceptible to a dilute gas
analysis, we have used a nonperturbative double Gaussian wave
function approximation in the quantum mechanical problem for
determining the eigenvalues of the transfer operator. The study of the
intermediate temperature regime is important because kink
contributions to some thermodynamic quantities (e.g., the specific
heat) of the system may be dominant precisely in this region. Numerical
calculations of the partition function, while certainly valuable,  are
by themselves not sufficient. Such calculations, for example, cannot
explain the nature of the peak in the specific heat nor can they
enable one to extract which effects are due to kinks and which effects
due to phonons.

Stated in brief, our results are: (1) the dilute gas predictions for
the kink density and the correlation length are very accurate below a
certain (theoretically estimable) temperature, (2) above this
temperature the Gaussian results for the kink number and correlation
length agree with the simulations, (3) kinks are found to
``disappear'' into the thermal phonon background above a
characteristic temperature, in good agreement with our theoretical
prediction, (4) our Gaussian approximation accurately describes the
classical single point field distribution function at high and
intermediate temperatures where the dilute gas (WKB) approximation
breaks down, (5) the internal energy and the specific heat calculated
in double Gaussian approximation show an interesting energy sharing
process between kinks and nonlinear phonons in an intermediate
temperature range below the characteristic temperature at which kinks
appear: a peak in the specific heat in this temperature range is shown
to be due essentially to kinks.

The rest of this paper is organized as follows. We describe our
computer simulation and numerical techniques in Sec. II. Sec. III is a
brief review of the transfer operator formalism and the standard WKB
results for the partition function. Here we also introduce a new,
``effective'' definition of the kink number that takes into account an
averaging over the thermal phonon length scale. In Sec. IV we
introduce the double Gaussian variational method and use it to find
approximate eigenfunctions and eigenvalues of the transfer operator
and also to compute thermodynamic quantities such as the specific
heat. Finally, we conclude in Sec. V with a discussion of our results
and of directions for future work.
\newpage

\centerline{\bf II. The Simulations}
The Langevin equation for the dimensionless theory is
\bq
\pa^2_{\bar{t}\bar{t}}\phi=\pa^2_{\bar{x}\bar{x}}\phi-
\eta\pa_{\bar{t}}\phi+\phi(1-\phi^2) + F(\bar{x},\bar{t}). \label{le}
\eq
To guarantee an approach to equilibrium, the Gaussian, white noise $F$
and the viscosity $\eta$ are related via the fluctuation-dissipation
theorem:
\bq
\VEV{F(\bar{x},\bar{t}) F(\bar{y},\bar{s})} = 2\eta\bar{\b}^{-1}\d
(\bar{x}-\bar{y}) \d (\bar{t}-\bar{s}).
\eq
Since in this paper we are interested only in the equilibrium
properties of the system which are independent of the viscosity, we
fixed $\eta=1$ for all the simulations. (It was verified that the
results did not depend on the value of $\eta$.)

We carried out numerical simulations of this Langevin equation using a
standard first order Euler differencing technique. The second order in
time Langevin equation (\ref{le}) was written as the two first order
finite difference equations
\ba
v(\bar{t}+\e)&=&v(\bar{t})+\e\left[\left({\bar{\f}(\bar{x}+\d)+
\bar{\f}(\bar{x}-\d)-2\bar{\f}(\bar{x})}\right)/\d^2\right.\nonumber\\
&&\left.-\eta v(\bar{t})+\bar{\f}(\bar{x})(1-\bar{\f}^2(\bar{x}))+
F(\bar{x},\bar{t})\right]
\nonumber\\
\bar{\f}(\bar{t}+\e)&=&\bar{\f}(\bar{t})+\e v(\bar{t}+\e)  \label{step}
\ea
where the ``velocity,'' $v=\pa_{\bar{t}}\bar{\f}$. The finite
differencing was implemented with a time step $\e=0.02$, and a
lattice spacing $\d = 0.5$. The spacetime noise was generated by
summing $N$ random variables uniformly distributed over $(-.5,.5)$.
{}From the central limit theorem, in the limit that $N \to \infty$ this
sum should approach a Gaussian random variable with mean $0$ and
variance $\sqrt{N/12}$. For convenience we chose $N=12$: since this
gave a noise normalized to a strength of unity, for each given
temperature the normalization was straightforwardly determined by the
fluctuation-dissipation relation. The results presented here were
obtained on lattices consisting of 16384 sites and we have checked their
consistency with results from lattices of different sizes. The lattice
volume is large enough that there are no discernible finite size effects.

Our system size is one to two orders of magnitude larger than that in
most previous simulations. Large system sizes are necessary to get
acceptable statistics at low temperatures. In many of the recent
numerical simulations of this model, the system size was such that
only a few kinks would appear at the lowest temperatures studied.
For temperatures in the range where the WKB theory is valid, fewer
than one kink would appear on systems of the sizes used in Refs.
\cite{GR}\cite{BD}\cite{AFG}.

The lattice systems were evolved from random initial configurations
to equilibrium. The length of time necessary to ensure equilibrium
increased with inverse temperature. For $\bar{\beta} = 8$ the time
required was approximately $10^7$ time steps, and for the highest
temperatures, less than $10^5$ steps.

Two quantities of interest reported here are the kink number and the
field correlation length. To compute the kink number, we need an
operational way to identify kinks, even though there is an exact kink
solution available theoretically. We therefore examined several
possible definitions, all of which rely on a knowledge of the
canonical kink size.  From the classical solution for a kink centered
at $x_0$, $\phi=tanh((x-x_0)/\sqrt{2})$, the kink scale $L_k$ is
approximately 8 lattice units.  Raw kink configurations are shown in
Fig. I. At low enough temperatures ($\bar{\beta} >5$), kinks may be
identified easily, however at higher temperatures this is clearly not
the case.

The naive approach is to simply count the number of zero-crossings of
the field, since one may argue that these are the ``tunneling events''
which correspond to kinks. However, at higher temperatures there are
zero-crossings due to thermal noise (phonons), and counting all
zero-crossings would lead to a gross overestimation of the number of
kinks. At high temperatures it is not possible to distinguish
unambiguously between kinks and nonlinear phonons. However, the
problem of the possible overcounting arises already at intermediate
temperatures, where kinks are distinct. A possible solution is to use
a smoothed field by either averaging or ``block-spinning'' the actual
field configuration over a length of the order of the kink length
scale. The latter approach was taken in previous simulations
\cite{GR}\cite{BD}\cite{AFG}. This solution is not without flaws
either, as rapid fluctuations can still appear as kinks. We prefer to
count kinks in the following way: at a particular time we first find
all zero-crossings. To test the legitimacy of a given zero-crossing we
check for zero-crossings one kink scale (8 lattice units) to its right
and to its left. If no zero-crossings are found, we count it as a
kink, otherwise not.

The number of kinks is plotted against $\bar{\beta}$ in Fig. II. Above
$\bar{\b}\sim 6$, the averaged field method and our method for counting
kinks agree. Moreover, in this (low temperature) range, the dilute gas
expression for the kink number (\ref{11}) is in excellent agreement
with the data. At elevated temperatures, there is a clear disagreement
between the two methods of counting kinks. The average field technique
has the number of kinks monotonically increasing with temperature;
whereas, in accord with intuition and the behavior of $P[{\phi}]$, the
second technique clearly shows a reduction in the kink number at
higher temperatures. (See Sec. IV for a detailed discussion.)
Moreover, in this temperature regime the number of kinks computed with
the smoothing method depends strongly on the smoothing scale. We
conclude that for $\bar{\b} < 6$, an unambiguous number of kinks cannot
be extracted with any confidence from the smoothing method.
Unfortunately, this is precisely the temperature regime explored in
previous simulations. We will discuss an analytic definition of the
kink number $N_k$ in Sec. III and show that this quantity can be
independent of the smoothing scale only at low temperatures.

The correlation length was extracted from the field configurations by
taking the inverse Fourier transform of the power spectrum and then
fitting an exponential decay to the correlation function. The
correlation length $\l$ is plotted against $\bar{\beta}$ in Fig. III.
The WKB prediction is seen to hold for $\bar{\b}>6$ while at higher
temperatures ($\bar{\b}>4$), the double Gaussian approximation is in
excellent agreement with the data.

Our numerical results can be understood by using different analytical
approaches in different temperature regimes. At low temperatures
($\bar{\b}>6$) the WKB method turns out to be very accurate. At higher
temperatures a double Gaussian variational method provides a good
description of our data. Both analytical methods are conveniently
discussed in the framework of the transfer operator formalism, which
we describe in the following section.
\newpage

\centerline{\bf III. The Transfer Operator}

The key idea behind the transfer operator method is to transform the
problem of finding the canonical partition function for a system to the
exactly equivalent problem of finding the eigenvalues of a certain
integral operator. With some smoothness assumptions for the
eigenfunctions of this transfer operator, it is then possible to show
that the problem reduces to that of finding the energy eigenvalues
of a related quantum mechanical problem. The method easily generalizes
to the problem of calculating correlation functions. The presentation
given here closely follows that of Ref. \cite{SSF}.

The canonical partition function for the Lagrangian (\ref{1}) is
given by the functional integral
\begin{eqnarray}
Z = \int D\Phi D\pi~\exp\left(-\b H[\Phi,\pi]\right) \label{Z}
\end{eqnarray}
where $\pi$ is the canonical momentum of the field and $H$ is the
field Hamiltonian. The integral over the momenta is a trivial Gaussian
integral. Writing
\bq
Z=Z_{\p}Z_{\F},                                \label{Zdef}
\eq
we have
\bq
Z_{\p}=\left({2\p\over\D\b}\right)^{N/2}    \label{Zp}
\eq
obtained by discretizing the spatial lattice with lattice spacing
$\D$. Periodic boundary conditions are assumed and the total
lattice size $L=N\D$. We will be primarily interested in obtaining the
configurational partition function, $Z_{\F}$. Skeletonizing the formal
path integral for $Z_{\F}$ by first forming
\bq
f(\F_{i+1},\F_i)=-{1\over 2}m^2\F_{i+1}^2+{1\over 4}\l\F_{i+1}^4
+{1\over 2}\left[\F_{i+1}-\F_i\over\D\right]^2     \label{f}
\eq
we have,
\bq
Z_{\F}=\prod_{i=1}^N\int d\bar{\F}_i~\exp\left(-\b\D
f(\F_{i+1},\F_i)\right)                                 \label{skel}
\eq
where as just stated we have imposed a periodic boundary condition
($\F_1=\F_{N+1}$), and
\bq
d\bar{\F}_i=\sqrt{\b\over 2\p\D}d\F_i.     \label{meas}
\eq
The normalization (\ref{meas}) is fixed by requiring equipartition to
hold for a free scalar field theory.

Introducing the transfer operator specified by the eigenvalue
problem
\bq
\int d\bar{\F}_i~\exp\left(-\b\D f(\F_{i+1},\F_i)\right)
\J_n({\F}_i) = \exp\left(-\b\D\e_n\right) \J_n({\F}_{i+1})
                                                   \label{to}
\eq
where $\e_n$ is the eigenvalue corresponding to the eigenfunction
$\J_n$. Substituting in (\ref{skel}), it follows from the completeness
of the eigenfunctions that
\bq
Z_{\F}=\sum_n\hbox{e}^{-\b L\e_n}.                   \label{toZ}
\eq
In the thermodynamic limit, $L\to\infty$, $Z_{\F}$ is determined
by the smallest eigenvalue of the transfer operator. In this limit,
the free energy density is simply $F_{\F}=\e_0$.

In an entirely analogous manner, it is possible to show that the
correlation function
\ba
\VEV{\F(x)\F(0)}&=&{1\over Z_{\F}}\int D\bar{\F}~\hbox{e}^{-\b
H[\F]}\F(x)\F(0)                         \nonumber\\
&=&\sum_n \abs{\bra{\J_n}\F\ket{\J_0}}^2\exp\left(-\b
\abs{x}\D_{0n}\right),                                  \label{corr}
\ea
where $\D_{0n}=\e_n-\e_0$. At large values of $x$, the lowest excited
state controls the behavior of the correlation function.

The problem is now to compute the eigenvalues and eigenfunctions of
the transfer operator. Assuming that the $\J_n({\F}_i)$ are smooth
functions of the $\F_i$, we Taylor expand $\J_n({\F}_i)$ on the
left hand side of (\ref{to}) in terms of the $\J_n({\F}_{i+1})$ on
the right hand side. The integral on the left hand side can now be
evaluated term by term. To leading order in $\D/\b$, (\ref{to})
becomes
\bq
\exp\left(-\b\D H_Q\right)\J_n=\exp\left(-\b\D\e_n\right)\J_n
                                            \label{sm}
\eq
which is equivalent to the time independent Schrodinger equation
\bq
H_Q\J_n=\e_n\J_n.                                  \label{se}
\eq
Dropping the indices on the fields,
\bq
H_Q=-{1\over 2\b^2}{\pa^2\over\pa\F^2}-{1\over 2}m^2\F^2+{1\over
4}\l\F^4+V_Q                           \label{ham1}
\eq
is the quantum Hamiltonian for a particle in a double well potential
with a temperature dependent energy shift
\bq
V_Q={1\over 2\b\D}\ln\left({\b\over 2\p\D}\right)    \label{vq}
\eq
arising from the normalization (\ref{meas}). The role of $\hbar$ is
now played by $\b^{-1}$, so that as the inverse temperature is made
larger, WKB becomes a better approximation. For the dimensionless form
of the theory (\ref{7}), the quantum Hamiltonian (\ref{ham1}) becomes
\bq
\bar{H}_Q=-{1\over 2\bar{\b}^2}{\pa^2\over\pa\f^2}-{1\over 2}
\f^2+{1\over 4}\f^4 + \bar{V}_Q                   \label{ham2}
\eq
where $\bar{V}_Q$ is just $V_Q$ with $\b$ replaced by $\bar{\b}$.
This is the form of the theory that we will use in the remainder of
the paper.

It is important to note that since (\ref{se}) was derived under
certain smoothness requirements, the eigenfunctions of $H_Q$ or of
$\bar{H}_Q$ will not always be eigenfunctions of the transfer
operator. This is the case at high temperatures where, for fixed $\D$,
the ratio $\D/\b$ may be much larger than unity. However, for the
intermediate and low temperature regimes which are of relevance to
kink thermodynamics, (\ref{se}) provides an adequate description. (If
needed, the high temperature behavior can be studied via perturbation
theory \cite{SSBB}.)

An alternative way to view (\ref{ham2}) is to introduce the scaled
field ${\bar{\f}}=\bar{\b}\f$, in which case,
\begin{equation}
\bar{H}_Q={1\over 2}\bar{p}^2 - {1\over 2\bar{\b}^2}{\bar{\phi}}^2 +
{1\over 4\bar{\b}^4}{\bar{\phi}}^4.             \label{ham3}
\end{equation}
Here we have omitted the contribution of $\bar{V}_Q$ which, since it
is just a shift, can be treated separately.

At low temperatures the two wells are widely separated and the ground
state energy is given by the oscillator ground state energy for one of
the wells minus the tunnel-splitting term, usually calculated by the
WKB method. One can then write
\bq
Z_{\f}=Z_{osc}Z_{tunn}Z_{\bar{V}_Q}                 \label{Zsplit}
\eq
or, in terms of the free energies,
\bq
F=F_{osc}+F_{tunn}+\bar{V}_Q                            \label{free}
\eq
where, from a crude WKB estimate,
\bq
F_{osc}\sim {L\over\sqrt{2}\bar{\b}}                 \label{Fosc}
\eq
and
\bq
F_{tunn}\sim -{L\over 2\sqrt{2}\bar{\b}}\hbox{e}^{-\bar{E}_k\bar{\b}}
\label{Ftunn}
\eq
where, the kink energy for the dimensionless form of the theory,
$\bar{E}_k=\sqrt{8/9}$. The key result of KS is the realization that
the tunneling contribution is associated with kinks and the sum of
{\em all} the other contributions with phonons \cite{KS}.

More generally one expects that there should be three qualitatively
distinct temperature ranges. At very low temperatures the
tunnel-splitting between two ``classically degenerate'' states is very
small and the WKB picture of kinks is valid. At higher temperatures
the wave function overlap in the region between the wells is large
enough that the WKB result for the energy difference between the
ground and first excited state is incorrect. One expects therefore
that for temperatures in this ``intermediate overlap'' regime, the WKB
approximation breaks down, but that kinks still exist as fairly well
identifiable objects. Moreover, the tunnel-splitting between the two
ground states can become comparable to the splitting between a ground
state and the first excited state inside each well (this, however, has
no effect on the correlation length, and as seen in Fig II, makes only
a minor correction to the kink number). Finally, at still higher
temperatures when the ground state energy in each well becomes larger
than the classical barrier height, the kink description fails since
the overlap is of order unity and the kink and phonon energy scales
become indistinguishable. These three temperature regimes are in fact
clearly seen in our numerical data (Fig. I) as well as in the
analytic results from the double Gaussian approximation of Sec. IV
(Fig. V).

Let us first describe the expectation values for the correlator and
the kink density based on the WKB approximation, which one expects to
be valid in the low temperature regime\footnote{From now on we perform
all calculations in the thermodynamic limit $L\rightarrow \infty$. The
large size of the system in our simulation (16384 sites) rules out
finite size effects: from (\ref{toZ}), it is clear that the
thermodynamic limit is correctly found in the simulation provided that
$L\D_{01}\bar{\b}\gg1$, or in terms of the correlation length, when
$L\gg\l$. At the lowest temperature studied here $L/\l>40$. Since this
is the worst case result, the condition $L\gg\l$ is very well
satisfied over the full range of the simulation.}. An accurate
calculation in this limit (see, e.g., Ref. \cite{CKBT}) yields,
for the energy difference between the ground and first excited state,
\bq
\D_{01}=4\left({3\bar{E}_k\over\bar{\b}\p}\right)^{1/2}
\exp(-\bar{E}_k\bar{\b}).  \label{WKBdiff}
\eq
The WKB result for the correlation function (using only the first two
states) is
\bq
\VEV{\f (0)\f (x)}\simeq\hbox{e}^{-\abs{x}/\l}  \label{lambda}
\eq
where the correlation length
\ba
\l&=&{1\over\bar{\b}\D_{01}}                   \label{cl}\\
&=&{1\over 4}\sqrt{\p\over
3}{1\over{\sqrt{\bar{E}_k\bar{\b}}}}
\exp\left(\bar{E}_k\bar{\b}\right). \label{lwkb}
\ea
This is plotted in Fig. III and is seen to fit the numerical data very
well for $\bar{\b}>6$.

Another quantity of interest is the number of kinks. In fact there
appears to be no unambiguous way to define this quantity. The number
of kinks is usually calculated in the phenomenological approach by
working in the grand canonical ensemble (see, e.g., Ref. \cite{CKBT}).
We now introduce a new way of defining an effective kink number by
working only with the original field variable. We begin by identifying
three length scales in the problem: $l_a$, the averaging scale, a
length scale long enough such that the averaged field
\bq
\f_a(\bar{x})={1\over l_a}\int_{-l_a/2}^{l_a/2}d\bar{y}~
\f(\bar{x}+\bar{y})     \label{avf}
\eq
is smooth on scales of the order of the kink length. Calling the
typical kink separation, $l_k$, and with $L_s$ such that
$l_a\le L_s<<l_k$, we define
\ba
N_k&=&{1\over 2}\VEV{{\left({\f}_a(0)-{\f}_a(L_s)\right)^2\over
4{\f}_0^{~2}}}                       \nonumber\\
&=&{1\over 4{\f}_0^{~2}}\left[\VEV{{\f}_a^{~2}}
-\VEV{{\f}_a(0){\f}_a(L_s)}\right]      \label{knum}
\ea
to be the kink number density. This definition comes from just
counting the number of zero-crossings of the smoothed field under an
assumption that the typical separation between the kinks is much
larger than $l_a$ and $L_s$. The normalization factor $\pm {\f}_0$ is
the asymptotic field value away from a kink. The angular brackets
denote a sampling over the whole lattice (divided into blocks of
length $L_s$). The factor $1/2$ compensates for the fact that
anti-kinks are also counted (the number of kinks is equal to the
number of anti-kinks). Note that our definition of kink number is
sensible only when there is an appropriate separation of length
scales, e.g., it is not valid at high temperatures when the kink
separation is very small (where even the notion of a kink is not well
defined). Even at intermediate temperatures, $N_k$ depends on the
smearing length. Such a dependence has indeed been noted in numerical
simulations \cite{KP}\cite{AH}.

To understand the errors in estimating the number of kinks from
(\ref{knum}) we first consider the case of $l_a$ fixed and varying
$L_s$. If, on the one hand, $L_s$ is chosen too large ($L_s>l_k$) then
there is an undercounting of the total number of kinks since there
will be many instances of having more than one kink in a block of
length $L_s$. If, on the other hand, $L_s$ is too small (i.e., smaller
than the kink size) then there will again be an undercounting since in
any given block of size $L_s$, $\abs{\f_a(0)-\f_a(L_s)}<2\f_0$.
Therefore, for a given $l_a$, one should maximize (\ref{knum}) with
respect to $L_s$. Now considering $l_a$ as variable, it is clear that
for small $l_a$ ($l_a$ less than the kink size), (\ref{knum}) can be
nonzero as long as nonnegligible phonon fluctuations are present {\em
independent} of whether kinks exist or not. At low temperatures one
expects $N_k$ to be independent of $l_a$ (see the discussion below)
but in the intermediate temperature regime there should be a
dependence on the averaging scale.

An explicit formula for $N_k$ can be obtained by substituting
(\ref{avf}) in (\ref{corr}). Then, for $L_s\ge l_a$, it is easy to
show that
\bq
N_k={1\over2\f_0^2\bar{\b}^2l_a^{~2}}
\sum_n{\abs{\bra{0}{\f}\ket{n}}^2\over\D_{0n}^2}
\left[\bar{\b}\D_{0n}l_a-1+\hbox{e}^{-\bar{\b}l_a\D_{0n}}+
\hbox{e}^{-\bar{\b}L_s\D_{0n}}
\left(1-\cosh(\bar{\b}\D_{0n}l_a)\right)\right].           \label{Nkm}
\eq
In the dilute gas regime we keep only the contribution of the first
excited state in (\ref{Nkm}), use the WKB result (\ref{WKBdiff}) for
$\D_{01}$, and take the limit $\bar{\b}\to\infty$ in (\ref{Nkm}). The
total number of kinks is, for $L_s\ge l_a$,
\ba
N_{tot}&\simeq&{L\over L_s}L_s\left(1-{l_a\over
3L_s}\right)\sqrt{3\over\p}\sqrt{\bar{E}_k\bar{\b}}
\hbox{e}^{-\bar{E}_k\bar{\b}}                   \nonumber\\
&=&L\left(1-{l_a\over
3L_s}\right)\sqrt{3\over\p}\sqrt{\bar{E}_k\bar{\b}}
\hbox{e}^{-\bar{E}_k\bar{\b}},                       \label{nkdg}
\ea
which, modulo the constant prefactor, is in agreement with the result
of Ref. \cite{CKBT} in this limit. An interesting point is that we
have an explicit expression for the prefactor and therefore can test
our formula directly against the simulations.

Our predictions are compared with results from the simulations in Fig.
II. (Note that the kink energy in the dimensionless form of the theory
$\bar{E}_k=\sqrt{8/9}$ and that $\bar{E}_k\bar{\b}=E_k\b$.) In the low
temperature limit, one expects the kink number to be independent of
$L_s$, and this is indeed true provided that $l_a\ll L_s$ (see
(\ref{nkdg})). In this case, while the analytic result overestimates
the number of kinks by a multiplicative factor of $\sim 1.6$, the
functional form of the temperature dependence of the kink number is in
very good agreement with the simulations. If one sets $L_s=l_a$ in
(\ref{nkdg}) then compared with the previous case, the kink number is
reduced by a factor of 2/3, and is in complete agreement with the
simulations. At present we do not understand the sensitivity of the
theoretical formula to the ratio of $l_a/L_s$ in the low temperature
regime. A discussion of the kink number density in the intermediate
temperature regime will be postponed to Sec. IV, after we have
described the double Gaussian approximation.

The agreement of the WKB/dilute gas results with the simulations imply
that there is no significant renormalization of the kink energy
(beyond that due to linearized phonons) at low temperatures (i.e.,
$\bar{\b} > 6$). This is in disagreement with the simulations
carried out in Ref. \cite{KBKS} but in agreement with the theory of
Ref. \cite{CKBT}. Whether there is or not such a renormalization of the
kink mass at intermediate temperatures is difficult to analyze as at
these temperatures the effects due to nonlinear phonons and
kinks are hard to disentangle. A good example of this is the behavior
of the kink number versus $\bar{\b}$ (Fig. II). Both numerical and
analytic results show that the ambiguity in the very notion of a kink
number density is such as to rule out any estimation of the kink mass
from the data at intermediate temperatures.

In the regime where WKB fails, one can compare the simulations of the
kink system with numerical solutions for the energy eigenvalues of the
Hamiltonian $\bar{H}_Q$. However, one would like to have a simple
analytical method for predicting the measured quantities. To this end
we implement the double Gaussian variational approximation in the
effective quantum mechanical problem. This approximation is an order
of magnitude more accurate than the simple Gaussian approximation for
this problem and correctly accounts for the reduction of energy due to
the overlap between the wave functions in the two wells. It also
allows a natural decomposition of phonon and kink contributions in
terms of separate ``diagonal'' and ``overlap'' contributions to the
ground state energy. The double Gaussian approximation should be
reliable at intermediate and high temperatures, where the overlap
between the two ground states is substantial (at low temperatures,
when the overlap is very sensitive to the form of the ``tail'' of the
wave function, WKB is more accurate).

\newpage

\centerline{\bf IV. The Double Gaussian Approximation}

The Gaussian approximation is a well-known nonperturbative variational
method for calculating the ground state energy and effective
potentials in quantum mechanics and quantum field theory \cite{PMS}. As
discussed in the previous section, the transfer operator technique
reduces the classical statistical mechanics of a field theory in $1+1$
dimensions to ordinary quantum mechanics. To apply the standard
Gaussian approximation is simple: one assumes that the ground state
wave function is a Gaussian with width $\O^{-1/2}$, and centered at
the point $\bar{\f}_0$,
\bq
\J_G(\bar{\f}_0,\O)=\sqrt{\O\over\p}\exp\left(-{1\over
2}\O(\bar{\f}-\bar{\f}_0)^2\right).
                                            \label{g1}
\eq
Next one computes the energy
\bq
V_G(\bar{\f}_0,\O)\equiv\bra{\J_G}H_Q\ket{\J_G},      \label{g2}
\eq
and minimizes with respect to $\O$ to yield the Gaussian effective
potential $V_G(\bar{\f}_0)$. The global minimum of $V_G(\bar{\f}_0)$
is the ground state energy as calculated in the Gaussian
approximation.

In our case, with the Hamiltonian given by (\ref{ham3}), it is easy to
show that
\bq
V_G(\bar{\f}_0,\O)={1\over 4}\O-{1\over
2\bar{\b}^2}\left(\bar{\f}_0^{~2}+{1\over 2\O}\right)+{1\over
4\bar{\b}^4}\left(\bar{\f}_0^{~4}+{3\over
4\O^2}+{3\bar{\f}_0^{~2}\over\O}\right).             \label{g3}
\eq
Minimization of (\ref{g3}) with respect to $\O$ yields the cubic
equation
\bq
\O^3+{\O\over\bar{\b}^2}\left(1-{3\over\bar{\b}^2}
\bar{\f}_0^{~2}\right)-{3\over 2\bar{\b}^4}=0,      \label{g4}
\eq
the largest positive root of which is of interest. Substitution of
(\ref{g4}) in (\ref{g3}) enables us to write
\bq
V_G(\bar{\f}_0)=-{1\over 2\bar{\b}^2}\bar{\f}_0^{~2}+{1\over
4\bar{\b}^4}\bar{\f}_0^{~4}+{1\over 2}\O-{3\over
16\bar{\b}^4}{1\over\O^2}.                               \label{g5}
\eq
Note that $\bar{\f}_0=0$ is always a local minimum of
$V_G(\bar{\f}_0)$. Above a certain temperature, it becomes the global
minimum and one has a behavior reminiscent of a first order phase
transition in the effective potential (Fig. IV). Numerically, this
transition occurs at $\bar{\b}=3.12$ and one might interpret this as
predicting a disappearance of kinks at this temperature. This is not
quite correct, however, as we will soon see.

A problem with the Gaussian approximation is that it does not account
for tunneling. We know intuitively that for the double well
Hamiltonian, at large well separation, the ground state should be
described by a superposition of two localized wave packets, one in
each well. These wave packets may be taken to be Gaussians. With such
a wave function there will be cross-terms in $\bra{\J}H_Q\ket{\J}$
arising from overlaps between the individual Gaussians. Correctly
accounting for these overlap terms is clearly essential for our
problem.

We therefore modify the trial wave function by taking it as a
superposition of two Gaussians:
\bq
\J_{DG}={1\over
N}\left[\J_G(\bar{\f}_0,\O)+\J_G(-\bar{\f}_0,\O)\right]    \label{g6}
\eq
where the normalization factor
\bq
N^2=2\left[1+\hbox{e}^{-\O\bar{\f}_0^{~2}}\right].          \label{g7}
\eq
It is then a simple matter to show that
\ba
V_{DG}(\bar{\f}_0,\O)&=&\bra{\J_{DG}}\bar{H}_Q\ket{\J_{DG}}
\nonumber\\
&=&{2\over
N^2}\left[V_G(\bar{\f}_0,\O)+\hbox{e}^{-\O\bar{\f}_0^{~2}}
\left(-{\O^2\bar{\f}_0^{~2}\over 2}- {1\over 4\bar{\b}^2\O}+{3\over
16\bar{\b}^4\O^2} +{\O\over 4}\right)\right].               \label{g8}
\ea
The first term represents the ``diagonal'' contribution to the energy
while the rest are cross-terms from the overlap or ``off-diagonal''
contribution. In the framework of the phonon-kink model these terms
have a natural interpretation: the ``diagonal'' term corresponds to
a contribution to the free energy due to phonons (including anharmonic
corrections) while the overlap term is the kink contribution. In the
low temperature limit this reproduces the WKB splitting of the free
energy into phonons and kinks.

In principle, one can minimize $V_{DG}$ by following the same
procedure as for $V_G$. Unfortunately, this yields a transcendental
equation for $\O$ which cannot be handled analytically. A way out is
to first minimize (\ref{g8}) with respect to $\O$ {\em ignoring the
overlap terms completely}: this amounts to keeping $\O$ as given by
(\ref{g4}) in (\ref{g8}). The minimization with respect to
$\bar{\f}_0$ is then carried out by differentiating (\ref{g8}) with
respect to $\bar{\f}_0$ and setting the result to zero. While this
procedure appears to be rather a drastic simplification, it is still a
major improvement over $V_G$ both qualitatively and quantitatively.
First, $V_{DG}(\bar{\f}_0)$ always has a local {\em maximum} at
$\bar{\f}_0=0$ so that there is no abrupt change in the behavior of
$V_{DG}(\bar{\f}_0)$ with temperature, consistent with the fact that
there is no finite temperature phase transition in this model
($V_{DG}$ and $V_G$ are compared in Fig. IV). Second, it is an order
of magnitude more accurate than $V_G$ in estimating the ground state
energy (worst case error $\sim 1\%$ compared to $\sim 15\%$). Third,
and very importantly, we can now estimate the energy of the higher
excited states.

To estimate the energy of the first excited state we simply minimize
$V_{DG}^{(1)}\equiv\bra{\J_1}H_Q\ket{\J_1}$ where $\J_1$ is the
antisymmetric partner of $\J_{DG}$:
\bq
\J_1={1\over\bar{N}}\left[\J_G(\bar{\f}_1,\O_1)-
\J_G(-\bar{\f}_1,\O_1)\right]
                                                    \label{g9}
\eq
with $\bar{N}^2=2[1-\exp(-\O_1\bar{\f}_1^{~2})]$. It is to be stressed
that the minimization of the energy with respect to $\J_1$ has to be
done {\em independently} from that with $\J_{DG}$. We find that
\bq
V_{DG}^{(1)}(\bar{\f}_1,\O_1)={2\over
\bar{N}^2}\left[V_G(\bar{\f}_1,\O_1)-\hbox{e}^{-\O_1\bar{\f}_1^{~2}}
\left(-{\O_1^{~2}\bar{\f}_1^{~2}\over 2} - {1\over
4\bar{\b}^2\O_1}+{3\over 16\bar{\b}^4\O_1^{~2}} +{\O_1\over
4}\right)\right]              \label{g10}
\eq
The difference between the absolute minima of $V_{DG}^{(1)}$ and $V_{DG}$
is the tunnel-splitting energy $\D_{01}$ calculated in the double
Gaussian approximation. We now investigate the low temperature limit
of the double Gaussian approximation and contrast it with the WKB
result. First, from (\ref{g4}), it follows immediately that
at low temperatures,
\bq
\O\simeq {\sqrt{2}\over\bar{\b}}.    \label{g11}
\eq
Also, at low temperatures, $\bar{\f}_0\simeq\bar{\b}$. Substituting
these last two results in (\ref{g8}) and (\ref{g10}) we find,
\bq
V_{DG}\stackrel{\bar{\b}\rightarrow\infty}{\simeq}
{1\over\sqrt{2}\bar{\b}}+{3\over 32\bar{\b}^2}-{1\over
4}-{\hbox{e}^{-\sqrt{2}\bar{\b}}\over
4}\left[4-{1\over\sqrt{2}\bar{\b}}-{3\over 8\bar{\b}^2}\right]
\label{el}
\eq
and
\bq
V_{DG}^{(1)}\stackrel{\bar{\b}\rightarrow\infty}{\simeq}
{1\over\sqrt{2}\bar{\b}}+{3\over 32\bar{\b}^2}-{1\over
4}+{\hbox{e}^{-\sqrt{2}\bar{\b}}\over
4}\left[4-{1\over\sqrt{2}\bar{\b}}-{3\over 8\bar{\b}^2}\right].
\label{eh}
\eq
Polynomial corrections in powers of $\bar{\b}^{-1}$ to the leading
order result are due to anharmonic corrections while the exponential
terms represent the tunneling contribution. On subtracting (\ref{el})
from (\ref{eh}) the tunnel-splitting term follows immediately:
\bq
\D_{01}\stackrel{\bar{\b}\rightarrow\infty}{\simeq}
{\hbox{e}^{-\sqrt{2}\bar{\b}}\over
2}\left[4-{1\over\sqrt{2}\bar{\b}}-{3\over 8\bar{\b}^2}\right]
\eq
which may be contrasted with the WKB result (\ref{WKBdiff}). In this
regime, the WKB result is more accurate as the double Gaussian method
tends to underestimate the overlap contribution
($\D_{01}\sim\exp(-\sqrt{2}\bar{\b})$ in double Gaussian versus
$\D_{01}\sim\exp(-\sqrt{8/9}\bar{\b})$ in WKB). This is to be expected
since in this regime the overlap between the two wave packets is very
small and therefore very sensitive to the form of the tail of the wave
function. At higher temperatures, however, the overlap is not small
and the sensitivity to the form of the tail disappears. Consequently,
(\ref{g8}) and (\ref{g10}) become increasingly more accurate.

At low temperatures, the second excited state is one of a pair of
tunnel-split harmonic oscillator first excited states. The state with
even parity does not contribute to the correlation function or to the
approximate kink number formula because the matrix element of the
position operator between the ground state and any even state
vanishes. The odd state, written in our double Gaussian approximation
is,
\bq
\J_2={1\over\tilde{N}}\left[(\bar{\f}-\bar{\f}_2)\J_G(\bar{\f}_2,\O_2)+
(\bar{\f}+\bar{\f}_2)\J_G(-\bar{\f}_2,\O_2)\right],        \label{g12}
\eq
and
\bq
\tilde{N}^2={1\over\O_2}+\hbox{e}^{-\O_2\bar{\f}_2^{~2}}
\left({1\over\O_2}-2\bar{\f}_2^{~2}\right).
\eq
The expectation value of the energy in this state now follows from a
straightforward calculation:
\ba
V_{DG}^{(2)}&\equiv&\bra{\J_2}H_Q\ket{\J_2} \nonumber\\
&=&{3\over 4\tilde{N}^2}\left[1-{1\over\bar{\b}^2\O_2^{~2}}
\left(1+{2\over 3}\O_2\bar{\f}_2^{~2}\right)+
{1\over\bar{\b}^4\O_2^{~3}}\left({5\over
4}+3\O_2\bar{\f}_2^{~2}+{1\over 3}\O_2^{~2}\bar{\f}_2^{~4}
\right)\right. \nonumber\\
&&+\hbox{e}^{-\O_2\bar{\f}_2^{~2}}\left\{1-
4\O_2\bar{\f}_2^{~2}+{4\over 3}\O_2^{~2}\bar{\f}_2^{~4}-
{1\over\bar{\b}^2\O_2^{~2}}\left(1-{2\over
3}\O_2\bar{\f}_2^{~2}\right) \right. \nonumber\\
&&\left.\left.+{1\over\bar{\b}^4\O_2^{~3}}\left({5\over
4}-{1\over 2}\O_2\bar{\f}_2^{~2}\right)\right\}\right]. \label{V2t}
\ea
In order to determine the values of $\bar{\f}_2$ and $\O_2$ one has in
principle to carry out a constrained minimization of the energy by
also enforcing the requirement that $\J_1$ and $\J_2$ be orthogonal.
{}From (\ref{g9}) and (\ref{g12}) this condition turns out to be
\bq
{\bar{\f}_2-\bar{\f}_1\over\bar{\f}_2+\bar{\f}_1}=
\exp\left(-{2\O_1\O_2\over\O_1+\O_2}\bar{\f}_1\bar{\f}_2\right). \label{ov}
\eq
For the range of temperatures we are interested in, the above condition
is reasonably well satisfied if we take $\O_1=\O_2$ and
$\bar{\f}_1=\bar{\f}_2$, i.e., the left hand side of (\ref{ov}) is
identically zero, while the right hand side is small (worst case of
order $10^{-2}$). Therefore for this state we do not need to carry out
a minimization of the energy as long as high accuracy is not required.
In the low temperature limit,
\bq
V_{DG}^{(2)}\stackrel{\bar{\b}\rightarrow\infty}{\simeq}
{3\over\sqrt{2}\bar{\b}}+{15\over 32\bar{\b}^2}-{1\over
4}+O({\hbox{e}^{-\sqrt{2}\bar{\b}}}.)              \label{V3}
\eq
The first term in (\ref{V3}) is just the energy of the first excited
harmonic oscillator state as expected. The other terms represent
anharmonic corrections and overlap contributions as usual. In Fig. V,
the first three energies (as calculated in double Gaussian
approximation) are plotted against the inverse temperature. While at
intermediate temperatures $\D_{02}$ and $\D_{01}$ are of similar
magnitude, at low temperatures, $\D_{02}\gg\D_{01}$ which of course is
just the WKB regime. It is clear that for values greater than
$\bar{\b}\sim 6$, the WKB results should hold.

The behavior of the energies of the ground and first two excited
states (Fig. V) allows the identification of three qualitatively
different regimes: (1) all the energies lie above the classical
barrier, (2) the ground state energy lies below the classical barrier
height (at $\bar{\b}=1.734$), and (3) the energy difference between
the first two states becomes negligible in comparison with the energy
difference between the ground and the second excited state (at
$\bar{\b}\sim 6$). Our simulations confirm the theoretical
expectations of the previous section that kinks cannot be identified
in region (1), that there are kinks, but that the dilute gas
approximation is invalid in region (2), and finally, that the dilute
gas approximation is accurate in region (3).

The classical single point field distribution function $P[\bar{\f}]$
can be measured directly from our simulations. For the analogous quantum
mechanical problem arising from the transfer operator method this is
just the square of the ground state wave function $\Psi_0$. Results
from the simulations are compared with $\J_{DG}^2$ at $\bar{\b}=2$ and
$\bar{\b}=4$ in Fig. VI and are in reasonable agreement. The presence
of kinks implies a double peak in $P[\bar{\f}]$ \cite{ADB} (the
converse is false) while a single peak centered at the origin means
that kinks and large amplitude thermal phonons can no longer be
distinguished. From the simulations such a transition occurs at
$\bar{\b}\simeq 1.7$, in agreement with the theoretical calculation of
when $\Psi_{DG}^{~2}$ goes over from a double to single peaked
distribution. (The double Gaussian method compares favorably with the
numerical evaluation of the transfer operator in Ref. \cite{SSBB},
which predicts $\bar{\b}\simeq 1.8$ as the transition point.)  As
expected, this is also the temperature (see Fig. 2) where the ground
state energy crosses the classical barrier height. (A discussion of
various methods to determine this characteristic temperature is given
in Ref. \cite{ARB}.)

It is also apparent from Fig. VI that the double peaks in the
distribution function move inward from the minimum of the classical
potential as the temperature increases (eventually coalescing at
$\bar{\b}\sim 1.7$). Physically this can be understood as nonlinear
phonon corrections due to the fact that near each minimum, the
potential is not symmetric under reflection around the minimum.

The correlation function as determined by (\ref{corr}) can of course
be directly evaluated in the double Gaussian approximation. The
correlation length is determined by the long distance behavior of the
correlation function, and as is clear from (\ref{cl}), one needs only
the energy difference between the ground and first excited state to
determine the correlation length. In Fig. III the correlation length as
given by the double Gaussian and WKB approximations (see (\ref{13})) is
compared with the results from the simulations. As expected, the
double Gaussian prediction is borne out at intermediate temperatures
(upto $\bar{\b} \sim 4$) whereas the WKB results are in agreement with
the data at low temperatures (above $\bar{\b}\sim 6$).

The kink number as determined by (\ref{Nkm}) requires the calculation
of the matrix elements $\bra{0}\f\ket{n}$. The first two nonzero
matrix elements as given by the double Gaussian approximation are
\ba
\bra{0}\bar{\f}\ket{1}&=&\G\left[a_+\exp
\left({(\O\bar{\f}_0+\O_1\bar{\f}_1)^2\over
4\s}\right)-a_-\exp\left({(\O\bar{\f}_0-\O_1\bar{\f}_1)^2\over
4\s}\right)\right]                                \label{ovf1}\\
\bra{0}\bar{\f}\ket{2}&=&\G\left[({1\over
2\s}+a_+^2-a_+\bar{\f}_1)\exp\left({(\O\bar{\f}_0+\O_1
\bar{\f}_1)^2\over 4\s}\right)\right.      \nonumber\\
&&+\left.({1\over 2\s}+a_-^2+a_-\bar{\f}_1)\exp\left({(\O\bar{\f}_0
-\O_1\bar{\f}_1)^2\over 4\s}\right)\right],           \label{ovf2}
\ea
where
\ba
\G&=&{2\over\bar{N}\tilde{N}}\left({\O\O_1\over\s^2}\right)^{1/4}
\hbox{e}^{-{1\over2}(\O\bar{\f}_0^{~2}+\O_1\bar{\f}_1^{~2})} \label{G}\\
\s&=&{1\over 2}(\O+\O_1)         \label{sig}\\
a_{\pm}&=&{\O\bar{\f}_0\pm\O_1\bar{\f}_1\over \O+\O_1}. \label{a}
\ea
As the energies for the first three states are already known we can
now evaluate the kink densities as given by (\ref{Nkm}). Theoretical
estimates (for different choices of $l_a$ and $L_s$) are compared with
results from the simulation in Fig. II. The double Gaussian results
for the kink density are entirely consistent with the numerical data
at $\bar{\b}<5.5$ (beyond this temperature the double Gaussian
approximation underestimates the number of kinks due to the
underestimation of $\D_{01}$). Due to the ambiguity in the concept of
kink number discussed earlier, the significance of this comparison at
high temperatures is not clear.

The internal energy and the specific heat can be computed
straightforwardly from the double Gaussian approximation. The behavior
of the specific heat with temperature is of particular interest: it is
known that at a temperature relatively close to, but somewhat below
the characteristic temperature where kinks appear, the specific heat
attains a maximum value. Though a real phase transition cannot occur
in this system, it can be argued that this peak is a signal for the
emergence of a new degree of freedom, in this case presumably, the
kinks. While this peak has been seen in numerical calculations of the
partition function \cite{SSBB}, it has not been unambiguously
established that this feature is due to kinks. As remarked earlier,
the ground state energy as given by the double Gaussian approximation
breaks into two contributions: one given by an overlap contribution
and arguably associated with kinks, the other associated with phonons.
This feature allows one to analyze the nature of the peak in the
specific heat.

To begin, we recall the definitions,
\bq
U={\pa\over\pa\bar{\b}}\left(\bar{\b}F\right)   \label{inte}
\eq
and
\bq
C_v=-\bar{\b}^2{\pa^2\over\pa\bar{\b}^2}\left(\bar{\b}F\right)
\label{cv}
\eq
for the internal energy $U$ and the specific heat $C_v$. In our case,
the free energy $F$ has four contributions: one from the kinetic term
in the original field Hamiltonian, one from the term given by the
normalization of the functional integral, one from the overlap
contribution, and one from the (potentially nonlinear) oscillations
around the potential minimum of $\bar{H}_Q$. The first contribution
stems from (\ref{Zp}) giving rise to the free energy density
\ba
F_{\p}&=&-{1\over\bar{\b}L}\ln(Z_{\p})  \nonumber\\
&=&-{1\over 2\bar{\b}\D}\ln\left({2\p\over\D\bar{\b}}\right).
\label{f1}
\ea
The other pieces come from (\ref{Zsplit}). The contribution
$F_{osc}$ is just the first term of (\ref{g8}) while $F_{tunn}$ is the
second term. The normalization contributes the $\bar{V}_Q$ term. From
(\ref{f1}) and (\ref{vq}) we can immediately calculate the
contribution to the internal energy density and the specific heat (per
unit length) of the kinetic and measure terms. This gives just the
results for a free field theory:
\ba
U_0&=&{1\over\D\bar{\b}}     \label{ufree}\\
C_v^{(0)}&=&{1\over\D}.      \label{cvfree}
\ea
Put another way, (\ref{ufree}) and (\ref{cvfree}) are the
contributions from linear phonons. The nontrivial contributions are
therefore included exclusively in $F_{osc}$ and $F_{tunn}$. At low
temperatures, the WKB result for the ground state energy may be used
to show that \cite{CKBT}
\bq
C_v^{(\f)}\stackrel{\bar{\b}\rightarrow\infty}{\simeq}
2\sqrt{{3E_k\bar{\b}\over\p}}\left[\left({1\over
2}-E_k\bar{\b}\right)^2-{1\over 2}\right]\hbox{e}^{-E_k\bar{\b}}.
\label{lt}
\eq
In this regime $F_{osc}$ does not contribute to the specific heat at
all while the kink contribution is exponentially suppressed. The
constant term due to linear phonons (\ref{cvfree}) is dominant.

The situation can be dramatically different at intermediate
temperatures. The specific heat as calculated from the double Gaussian
approximation is plotted in Fig. VII: A prominent peak in the specific
heat appears at $\bar{\b}\simeq 5.4$. To understand whether this peak
is due to kinks, the individual contributions from $F_{osc}$ and
$F_{tunn}$ are also plotted. There seems to be a delicate interplay
between these two contributions. While the rise to the peak with
increasing $\bar{\b}$ is due to nonlinear phonons, just before the
peak this contribution drops off steeply and eventually leads to a
{\em reduction} in the height of the peak. The overlap or kink
contribution exhibits a slow initial decline as $\bar{\b}$ increases,
followed by a relatively sharp peak: it is this peak that is the
dominant contribution at $\bar{\b}\sim 5.4$. We can therefore conclude
that it is indeed the kinks that are responsible for the peak in the
specific heat at this temperature. At larger values of $\bar{\b}$,
both contributions fall off smoothly to zero as expected from the low
temperature result (\ref{lt}).

We have also calculated the internal energy $U$. The nonlinear phonon
and kink energies are plotted in Fig. VIII. There is a clear indication
of an intricate energy sharing mechanism operating between the
nonlinear phonons and kinks at intermediate temperatures with kinks
dominating the internal energy in the neighborhood of $\bar{\b}\sim
5.4$. Presumably this is due to the fact that at these temperatures
kinks emerge as well defined localized objects and are created with
relative ease.

The results from the double Gaussian approximation as shown in Figs.
VII and VIII are not quantitatively trustworthy for $\bar{\b}>5$ as
the approximation breaks down at that point. However, we expect them
to be correct qualitatively: since this method underestimates the
overlap at low temperatures, it in fact suppresses the kink
contribution at lower temperatures. Therefore it is highly unlikely
that the peak in the specific heat due to the kink contribution is an
artifact of the approximation. A direct comparison of the theoretical
results for the specific heat with numerical simulations is possible
but requires high statistics. Work in this direction is in progress.

\newpage
\centerline{\bf V. Discussion}

In summary, we have shown that the dilute gas/WKB approximation is
excellent for $\bar{\b}> 6$ with no further phonon dressing of the bare
kink energy beyond that already included in (\ref{11}) and (\ref{13})
at these low temperatures. In particular, we see no evidence of a
reduction in the kink mass as suggested in some previous simulations
\cite{BD}\cite{AFG}. At higher temperatures, the WKB analysis
fails, though theoretical progress is possible with the double
Gaussian technique. In this temperature regime, we have shown that the
correlation length as measured in the simulations is in very good
agreement with the theory and that the kink number density, while not
being a well defined quantity, can still be understood at least
qualitatively. The double Gaussian approximation also accurately
predicts the onset of short-range order in the system as evidenced by
the probability distribution function crossing over from a single to a
double peaked distribution.

The work reported here contains some of the largest simulations
carried out to date on the $\F^4$ model. The advantage of size is
apparent: we have been able to go to low enough temperatures to
unambiguously verify the WKB predictions for this model. Furthermore,
the large system size enabled us to check for finite size effects and
to be confident of their absence.

The decomposition of the specific heat into two contributions via the
double Gaussian approximation appears to demonstrate the existence of
a nontrivial energy sharing interaction between kinks and nonlinear
phonons in the neighborhood of the temperature where the specific heat
is a maximum. At this temperature, $(\bar{\b}\sim 5.4)$, kinks exist as
well-defined objects (Fig. I) but the dilute gas/WKB theory is not valid.
It would be interesting to see if an analytic study of the
thermodynamics of kinks and phonons is possible in a phenomenological
theory in this temperature regime.

We expect to apply the double Gaussian variational method to other
problems in the future, e.g., Sine-Gordon and $\F^6$ field theories in
$1+1$ dimensions. Work on the Langevin simulation method applied to
calculate dynamical correlation functions is already in progress.

\centerline{\bf Acknowledgments}

We thank M. Alford, S. Chen, F. Cooper, G. D. Doolen, H. Feldman, R.
Gupta, R. Mainieri, M. Mattis, E. Mottola, A. Saxena, W. H. Zurek, and
especially, A. R. Bishop, for encouragement and helpful discussions.
This work was supported by the U. S. Department of Energy at Los
Alamos National Laboratory and by the Air Force Office of Scientific
Research. Numerical simulations were performed on the CM-200 at the
Advanced Computing Laboratory at Los Alamos National Laboratory and
the CM-2 at the Northeast Parallel Achitecture Center at Syracuse
University.

\newpage

\centerline{\bf Figure Captions}

\noindent {\bf Fig. I:} Sample field configurations, from top to
bottom, at $\bar{\b}=2$, $\bar{\b}=4$, $\bar{\b}=5.5$, and
$\bar{\b}=8$. Only a 1000 lattice unit sample of the total lattice
size of 16384 is shown.

\noindent {\bf Fig. II:} Total number of kinks as a function of
$\bar{\b}$. Squares denote counts with a smoothed field ($l_a=8$
lattice units) definition of kinks, diamonds for the zero-crossing
counting method discussed in Sec. II, and the solid line starting at
$\bar{\b}=4$ is a fit to the WKB prediction (\ref{11}). Also shown are
three predictions (the three pairs of solid lines) for the kink number
from the theoretical formula (\ref{Nkm}) calculated in double Gaussian
approximation. Three values for the averaging length were used, from
top to bottom, $l_a$=2,4, and 8 lattice units. For each value of $l_a$,
there are two theoretical curves: (1) calculating (\ref{Nkm}) keeping
only the ground and first excited states and (2) keeping the ground,
and first and second excited states (the upper curves of each pair).
The kink number is seen to strongly depend on $l_a$ for $\bar{\b}<6$.
(The double Gaussian approximation breaks down at higher values of
$\bar{\b}$.)

\noindent {\bf Fig. III:} Field correlation length $\l$ as a function
of $\bar{\b}$. The double Gaussian (top curve) and WKB (bottom curve)
predictions are compared with the numerical results. Squares are data
points from the simulations. The crossover from the double Gaussian to
the WKB range of validity happens around $\bar{\b}\sim 5$.

\noindent {\bf Fig. IV:} The Gaussian effective potential $V_G$
(dashed lines) compared with $V_{DG}$ (dot-dashed lines) at two
temperatures ($\bar{\b}=2.5$ and $\bar{\b}=4$). The classical
potential $V=-(1-\bar{\f}^2/(2\bar{\b}^2)) \bar{\f}^2/(2\bar{\b}^2)$
from (\ref{ham3}) is also shown (solid curves). For $\bar{\b}<3.12$,
$V_G$ has a global minimum at the origin, while for $\bar{\b}<3.12$
there are two degenerate global minima. For $V_{DG}$ the
origin is always a maximum and there are always two degenerate minima.
The ground state energy calculated in the two approximations is the
global minimum of $V_G$ or of $V_{DG}$. The ground state energy is
always lower in the double Gaussian approximation.

\noindent {\bf Fig. V:} The ground state, first excited, and the odd
second excited state energies computed in double Gaussian
approximation (the lowest, middle and uppermost solid lines) plotted
against $\bar{\b}$. The ground state energy has further been
decomposed into the contributions from the ``diagonal'' (dot-dashed
line) and ``overlap'' (dashed line) pieces.

\noindent {\bf Fig. VI:} The classical distribution function
$P[\bar{\f}]$ (solid lines) given by the simulation and the
distribution $\J_0^{~2}$ from the double Gaussian approximation
(dashed lines) plotted against $\bar{\f}$ for $\bar{\b}=2$ and
$\bar{\b}=4$. The underestimation of the wavefunction overlap in the
double Gaussian approximation is already visible at $\bar{\b}=4$.

\noindent {\bf Fig. VII:} The specific heat $C_{v}$ as calculated from
the double Gaussian approximation plotted against $\bar{\b}$ (top
curve). The peak in the specific heat occurs at $\bar{\b}\simeq 5.4$.
The contribution of the linear phonons to the specific heat is the
constant value $1/\D=2$. The individual contributions arising from the
kinks (solid line) and nonlinear phonons (dot-dashed line) are shown
below. The WKB calculation for the kink contribution (dashed line) is
plotted from $\bar{\b}=6.5$ onwards.

\noindent{\bf Fig. VIII:} The internal energy plotted against
${\bar{\b}}$ in double Gaussian approximation (solid line). Kink
(dashed line) and nonlinear phonon (dot-dashed line) contributions are
shown separately (modulo an irrelevant constant energy shift). Note
that kinks dominate over nonlinear phonons in the region where the
specific heat has a maximum.

\end{document}